\documentstyle[epsfig,referee]{mn}

\newif\ifAMStwofonts

\ifoldfss
  \ifCUPmtlplainloaded \else
    \NewTextAlphabet{textbfit} {cmbxti10} {}
    \NewTextAlphabet{textbfss} {cmssbx10} {}
    \NewMathAlphabet{mathbfit} {cmbxti10} {} 
    \NewMathAlphabet{mathbfss} {cmssbx10} {} 
  \fi
  \ifAMStwofonts
    \ifCUPmtlplainloaded \else
      \NewSymbolFont{upmath} {eurm10}
      \NewSymbolFont{AMSa} {msam10}
      \NewMathSymbol{\upi}     {0}{upmath}{19}
      \NewMathSymbol{\umu}     {0}{upmath}{16}
      \NewMathSymbol{\upartial}{0}{upmath}{40}
      \NewMathSymbol{\leqslant}{3}{AMSa}{36}
      \NewMathSymbol{\geqslant}{3}{AMSa}{3E}

       \let\le=\leqslant
       \let\ge=\geqslant
    \fi
  \fi
\fi 

\ifnfssone
  \newmathalphabet{\mathit}
  \addtoversion{normal}{\mathit}{cmr}{m}{it}
  \addtoversion{bold}{\mathit}{cmr}{bx}{it}
  \newmathalphabet{\mathbfit} 
  \addtoversion{normal}{\mathbfit}{cmr}{bx}{it}
  \addtoversion{bold}{\mathbfit}{cmr}{bx}{it}
  \newmathalphabet{\mathbfss} 
  \addtoversion{normal}{\mathbfss}{cmss}{bx}{n}
  \addtoversion{bold}{\mathbfss}{cmss}{bx}{n}
  \ifAMStwofonts
    \ifCUPmtlplainloaded \else
      \UseAMStwoboldmath
      \makeatletter
      \new@mathgroup\upmath@group
      \define@mathgroup\mv@normal\upmath@group{eur}{m}{n}
      \define@mathgroup\mv@bold\upmath@group{eur}{b}{n}
      \edef\UPM{\hexnumber\upmath@group}
      \new@mathgroup\amsa@group
      \define@mathgroup\mv@normal\amsa@group{msa}{m}{n}
      \define@mathgroup\mv@bold\amsa@group{msa}{m}{n}
      \edef\AMSa{\hexnumber\amsa@group}
      \makeatother
      \mathchardef\upi="0\UPM19
      \mathchardef\umu="0\UPM16
      \mathchardef\upartial="0\UPM40
      \mathchardef\leqslant="3\AMSa36
      \mathchardef\geqslant="3\AMSa3E

       \let\le=\leqslant
       \let\ge=\geqslant
    \fi
  \fi
\fi 

\ifnfsstwo
  \DeclareMathAlphabet{\mathbfit}{OT1}{cmr}{bx}{it}
  \SetMathAlphabet\mathbfit{bold}{OT1}{cmr}{bx}{it}
  \DeclareMathAlphabet{\mathbfss}{OT1}{cmss}{bx}{n}
  \SetMathAlphabet\mathbfss{bold}{OT1}{cmss}{bx}{n}
  \ifAMStwofonts
    \ifCUPmtlplainloaded \else
      \DeclareSymbolFont{UPM}{U}{eur}{m}{n}
      \SetSymbolFont{UPM}{bold}{U}{eur}{b}{n}
      \DeclareSymbolFont{AMSa}{U}{msa}{m}{n}
      \DeclareMathSymbol{\upi}{0}{UPM}{"19}
      \DeclareMathSymbol{\umu}{0}{UPM}{"16}
      \DeclareMathSymbol{\upartial}{0}{UPM}{"40}
      \DeclareMathSymbol{\leqslant}{3}{AMSa}{"36}
      \DeclareMathSymbol{\geqslant}{3}{AMSa}{"3E}

       \let\le=\leqslant
       \let\ge=\geqslant
    \fi
  \fi
\fi 

\ifCUPmtlplainloaded \else
  \ifAMStwofonts \else 
    \def\upi{\pi}
    \def\umu{\mu}
    \def\upartial{\partial}
  \fi
\fi

\title{On formation rate of close binaries consisting of a super-massive black hole and
a white dwarf}
\author[P. B. Ivanov]
       {P. B. Ivanov \\
        Astro Space Center of PN Lebedev Physical 
Institute, Moscow, Russia
\\ Astronomy Unit, School of Mathematical Sciences, Queen Mary,
University of London, UK}

\pagerange{\pageref{firstpage}--\pageref{lastpage}}
\pubyear{1994}

\begin{document}

\maketitle

\label{firstpage}

\begin{abstract}
The formation rate of a close binary consisting of a super-massive black hole and
a compact object (presumably a white dwarf) in galactic cusps 
is calculated with help of the so-called loss cone approximation.
For a power-law cusp of radius $r_{a}$, the black hole mass $M\sim 10^{6} M_{\odot}$,
and the fraction of the compact objects $\delta \sim 0.1$
this rate $\dot N_{wd} \sim 4\cdot 10^{-5}K(p)\sqrt{{GM\over r_{a}^{3}}} 
\approx 3\cdot 10^{-9}K(p){({M\over 10^{6}M_{\odot}})}^{1/2}{({r_{a}\over 1pc})}^{-3/2}yr^{-1}$. The function 
$K(p)$ depends on parameter $p$ determining the cusp profile, and for the standard cusp
profiles with $p=1/4$ $K(p)\sim 2$. We estimate the  probability ${\it Pr}$ of finding of
a compact object orbiting around a black hole with the period $P$ in one particular galaxy 
to be
${\it Pr}\sim 10^{-7}{({P/10^{3}s\over M/10^{6}M_{\odot}})}^{8/3}{({M/ 10^{6}M_{\odot}\over r_{a}/ 1pc})}^{3/2}$. 
The object with the period
$P\sim 10^{3}s$  emits gravitational waves with amplitude sufficient to be 
detected
by LISA type gravitational wave antenna from the distance $\sim 10^{3} Mpc$. Based on
estimates of masses of super-massive black holes in nearby galaxies,
we speculate that LISA would detect several such events during its mission.    

\end{abstract}


\section{Introduction}

A compact object orbiting around a super-massive black hole with mass $\sim 10^{5}-10^{6}M_{\odot}$
with periods $\sim 10^{3}-10^{4}s$ produces gravitational radiation which could be detected
by future space-based gravitational antennas. As was mentioned by a number of authors, such
object could settle in a tight orbit around the black hole due to combined action of
two body gravitational encounters with other stars in galactic centers and emission of gravitational
radiation. The gravitational radiation coming from such object yields a direct information
about relativistic field of the black hole. Therefore, it is very interesting to estimate the rate
of production of close binaries consisting of a super-massive black hole and a compact object (the
capture rate below), and the probability ${\it Pr}$ of finding of a galaxy with such a binary
in the center. This problem has been investigated before by Hils and Bender 1995 and
Sigurdsson and Rees 1997. Hils and Bender performed numerical calculation of the capture rate
for a black hole and a central stellar cluster
with parameters similar to that was observed in the center of M32.
Sigurdsson and Rees (hereafter SR) generalized this result by making estimates 
of the capture rate for a more general model of the galactic centers. 
The purpose of this paper is to extend the results obtained before 
by calculation of the capture rate and the probability
${\it Pr}$ in frameworks of the so-called ``loss cone approximation'' (e. g. Lightman $\&$ Shapiro
1977, hereafter LS, and references therein). Also, we correct a mistake made in previous estimates.

In what follows we consider a black hole and a stellar cluster around it with parameters close 
to the parameters of the black hole and the stellar cluster in the center of our own Galaxy.  
At first, we assume that the black hole mass is about $10^{6}M_{\odot}$. Inside a radius where
the black hole mass is approximately equal to the total mass of the stars, a power law increase of
the star's number density with decreasing of distance (called the ``cusp'' in the stellar distribution)
is predicted by theory (e.g. Bahcall and Wolf 1976). Existence of the cusp is in agreement
with observations of the center of our Galaxy (e.g Alexander 1999). Based on observations
of our Galaxy and other nearby galaxies, we would expect the cusp radius of the order $\sim 1pc$.
As it will be clear from our discussion (see the next Section), in order to calculate the capture rate
we will be interested in inner regions of the cusps, where the density of normal stars is
strongly suppressed by star-star collisions.  Therefore, we will be interested in compact objects,
which could survive collisions with normal stars.
There is also another argument for paying special attention to the compact objects in our problem.
Namely, a normal star orbiting around a black hole
with period of interest must be tidally disrupted, and therefore only the compact objects with
small tidal radii may give a persistent source of gravitational radiation.  
As it was argued by SR,
it it reasonable to suppose that the galactic centers contain evolved population of stars,
with a large number of white dwarfs (the ratio of the total number of white dwarfs to
the total number of the normal stars $\delta \sim 0.1$). Since the capture rate is proportional
to the total number of compact objects, the white dwarfs could yield the main contribution
to the capture rate. Therefore, for simplicity, we consider below the innermost part of the
cusp consisting purely of white dwarfs. The generalization of our results on the case
of neutron stars and solar mass black holes is straightforward.

We calculate the capture rate in Section 2. In Section 3 we estimate the probability ${\it Pr}$.
We summarize and discuss our results in Section 4.

G is the Newton constant of gravity and c is the speed of light throughout the Paper.
We use expressions from  Gradshtein and Ryzhik 1994 when operating with special functions
without explicit referencing to that book.

\section{Capture rate}

For estimate of the capture rate we must compare influence of two
body gravitational encounters and emission of gravitational radiation
on the orbit of a white dwarf. At first we need a simple model of
distribution of normal stars and white dwarfs inside the galactic 
cusps. As a first approximation we use the standard power law isotropic
distribution function of stars in the phase space (e.g Spitzer 1987, 
Lightman $\&$ Shapiro 1977,
hereafter LS)
$$ f\approx CE^{p}, \eqno 1$$
where $C$ is a constant, $E=GM/r-v^{2}/2$ is the binding energy of a
star per unit of mass and $v$ is the velocity of a star. The parameter
$p < 3/2$ should be rather small from theoretical ground, 
it has two preferable values: $p=0$ (Peebles 1972, Young 1980) 
and $p=1/4$ (Bachall and Wolf 1976)
\footnote{ 
It is interesting to note that the same law $p=1/4$ was obtained by Gurevich 1964
in a similar problem of distribution of electrons around a charged body.}
.    
Estimates of this parameter from modeling of stellar distribution
in the center of our galaxy
made by Alexander 1999 also favor rather small
values. The number density of stars is obtained from (1) by
integration over the velocity space:
$$n(r)\approx n_{a}({r\over r_{a}})^{-(3/2+p)}, \eqno 2$$  
where $r_{a}$ is the radius of the cusp and $n_{a}$ is the number density
of stars at $r_{a}$. The constant $C$ is related to $n_{a}$ as
$$C=2^{-5/2}\pi^{-1}{({GM\over r_{a}})}^{-(p+3/2)}{n_{a}\over B(3/2;p+1)},
\eqno 3$$ 
where $M$ is the black hole mass and $B(x,y)$ is the Beta function
\footnote{ 
Note the useful relations: $B(x,y)=B(y,x)={\Gamma (x) \Gamma(y)\over
\Gamma (x+y)}$, where $\Gamma (x) $ is the  Gamma function.}  
The total mass of stars inside a sphere of radius $r < r_{a}$ is 
given as
$$ M(r)=4\pi\beta m n_{a}r^{3}_{a}{({r\over r_{a}})}^{1/\beta},
\eqno 4$$
where $m$ is the star's mass and 
$$\beta ={1\over 3/2-p}.$$ 
As we mentioned
in Introduction, the total mass of stars in the cusp must be of order of
the black hole mass $M$. In this paper we assume that these two masses
are equal: $M(r_{a})=M$. This gives the normalization condition:
$$n_{a}={M\over 4m \pi \beta r_{a}^{3}}. \eqno 5$$

We assign the subscript ``wd'' to all quantities describing white dwarfs.
It is assumed below that the distribution function of white dwarfs
$f_{wd}$ has the form analogous to the form of distribution (1), but
the total number of white dwarfs $N_{wd}$
is $\delta \sim 0.1$ times smaller than the total number of normal
stars $N=M/m$ (SR):
$$f_{wd}\approx \delta CE^{p} \eqno 6$$
All masses of white dwarfs are assumed to be equal
to the solar mass $M_{\odot}$, 
and their radii are equal to $0.01R_{\odot}
\approx 7\cdot  10^{8} cm$. 

Both distribution functions $f$ and $f_{wd}$ must be strongly modified
at sufficiently high energies, where star-star collisions come into play
and strongly reduce the number of stars with high binding energy. 
We would like to take this effect into account in
the simplest approximation possible. Therefore,
we assume that the distribution function
of normal stars is equal to zero at all energies exceeding the cutoff energy
$$E_{coll}\approx {Gm\over R_{star}}=(q{r_{a}\over R_{star}})
{GM\over r_{a}}, \eqno 7$$ 
where 
$$q=m/M\sim 10^{-6}.$$ 
Assuming that all normal stars have solar masses
and radii, we obtain another useful form of (7)
$$E_{coll}\approx 43 (q_{6}r_{1}){GM\over r_{a}}, \eqno 8$$
where $q_{6}=m/M_{6}$ and $M_{6}={M\over 10^{6}M_{\odot}}$, 
$r_{1}=r_{a}/1pc$. 
The cutoff energy of white dwarfs $E_{coll}^{wd}$ is approximately
one hundred times larger than $E_{coll}$:
$$E_{coll}^{wd}\approx 4300(q_{6}r_{1}){GM\over r_{a}}. \eqno 9$$

In our paper we are mainly interested in evolution of angular momentum
(per unit of mass) of low angular momentum orbits of stars due to
two body gravitational encounters. The change of angular momentum
per one orbital period is a random quantity,
and its dispersion $j^{2}_{2}$ plays a central role in our analysis.
This quantity can be calculated provided the distribution function of
stars in the phase space is given
(Chandrasekhar 1942, Rosenbluth at al 1957, Cohn $\&$ Kulsrud 1978, Spitzer 1987,
Bisnovatyi-Kogan et al and references therein)
as a function of orbital parameters of a star (i.e. its binding
energy $E$ and its angular momentum $J$). However, in general, it has
too complicated form. Therefore, we use several simplifying assumptions
when calculating this quantity. 1) The expression (2) is used for 
the number density of stars in the cusp. 2) $j^{2}_{2}$ is calculated     
in the low angular momentum limit ($J \rightarrow 0$). 3) We use additional
simplification for the distribution function 
of stars over velocity $F(v)$. Namely,
we take $F(v)$ in the form
$$F(v)=3v_{max}^{-3}v^{2}, \eqno 10$$
for $v < v_{max}$ and $F(v)=0$ for $v > v_{max}$. Here $v_{max}=
{({2GM\over r})}^{-3/2}$, and the factor $3v_{max}^{-3}$ in (10)
is determined by
the normalization condition $\int^{v_{max}}_{0}dv F(v)=1$. Equation (10)
is exact for $p=0$. Since we are mainly interested in rather low values
of $p$: $0 < p < 1/4$, the expression (10) is approximately valid for
the interesting range of $p$.
Even for values of $p \sim 1$, the expression (10) gives an error of order of
unity which is insignificant for our purposes
(see, however Cohn $\&$ Kulsrud 1978, Shapiro $\&$ Marchant 1978, Young 1980
who use the exact value of this quantity in numerical computations).   
Under these assumptions we have:
$$j^{2}_{2}={\kappa n_{a} r_{a}^{3/2+p}\over 10GM}{({GM\over E})}^{5/2-p}
I(p), \eqno 11$$
where $\kappa =8\pi (Gm)^{2}Ln \Lambda$, $Ln \Lambda \approx Ln (0.5N)$
is the standard Coulomb logarithm. The correction factor
$$I(p)=B(5/2-p; 1/2)\lbrace 4 +{5/2-p\over 3-p} \rbrace \eqno 12$$  
weakly depends on $p$: $I(0)=29\pi /6 \approx 5.65$ and $I(p=1/4)\approx
5.75$. 
It is instructive to compare the expression (11) with that was used by LS.
LS approximate $F(v)$ to be of the Maxwell form and obtain
$$j^{2}_{2}(LS)={0.501\over 2^{3/2}}\sqrt{(5+2p)}B(5/2-p;1/2){\kappa n_{a}r_{a}^{3/2+p}\over GM}
{({GM\over E})}^{5/2-p}.$$ The ratio $R=j^{2}_{2}(LS)/j^{2}_{2}$ has the following values:
$R(p=0)=0.82$, $R(p=1/4)=0.86$, $R(p=3/4)=0.95$ and $R(p=1)=0.99$. The deviation 
introduced by this difference in our final results is absolutely insignificant. 
However, as we have mentioned above, our expression is exact for $p=0$.

The expression (11) is valid for $E < E_{c}$. In the case
$E_{c} < E < E_{c_wd}$ this expression is $\delta$ times smaller:
$$j^{2}_{2_wd}=\delta j^{2}_{2}=C_{1}q(GMr_{a})\tilde E^{p-5/2}, \eqno 13$$
where the dimensionless energy variable
$$\tilde E={r_{a}E\over GM}, \eqno 14$$
and the correction factor
$$C_{1}={\delta Ln \Lambda I(p)\over 5\beta} \eqno 15$$
is of order of unity. We use equations (3), (5), (6) and (11) to obtain (13). 

For our purposes we  are interested in
the innermost part of the cusp, where the number density of normal
stars is suppressed due to star-star collisions.
Therefore, it is assumed that two body
gravitational deflections
of a test star are mainly produced by white dwarfs.
Generalization
of all equations on the case of deflection by normal stars is trivial.

It is very important to note that a black hole is able to capture
low angular momentum stars directly (see, e. g. Frolov $\&$ Novikov 1998
and references therein). Hereafter we would like to consider the simplest
case of Schwarzschild black hole where the process of direct capture can
be described in a very simple way. Namely, the star is captured by a
Schwarzschild black hole when its angular momentum $J$ is smaller than
the critical angular momentum: 
$$J_{crit}={4GM\over c} \eqno 16$$
which defines the size of ``loss cone'' in the stellar distribution.
Therefore, it is very convenient to use the dimensionless quantity
$$x={J\over J_{crit}}$$ 
instead of $J$ in our analysis. The corresponding
dispersion
$x^{2}_{2_wd}\equiv j^{2}_{2_wd}/J^{2}_{crit}$ can be
written as
$$x^{2}_{2_wd}={C_{1}q\over 4^{2}}({r_{a}c^{2}\over GM})
{\tilde E}^{p-5/2}. \eqno 17$$
In the simplest approximation influence of two body encounters
on the star's orbit can be described as follows.  
Let the dimensionless angular momentum of the test star $x$ to
walk randomly during some time interval $\Delta t$ due to two body 
gravitational encounters.
Then, the rms value of difference ${<\Delta x >}^{1/2}$ between 
initial and final values of $x$ is changing with time according
to the usual square root law:
$${<\Delta x^{2} >}^{1/2}\sim \sqrt {x^{2}_{2_wd}{\Delta t\over P(E)}}
\eqno 18$$
where 
$$P(E)={\pi GM\over \sqrt{2E^{3}}}=\pi\sqrt{{r_{a}^{3}\over 2GM}}
{\tilde E}^{-3/2} \eqno 19$$   
is the orbital period of the star.

Now let us consider influence of  emission 
of gravitational waves on the star's orbit. 
This influence is important for star's 
orbits with very high eccentricity $e \approx 1$. Therefore, we can 
use an approximate expression for the orbit decay time-scale $T{gw}$ 
obtained
by Peters 1964 in the limit of highly eccentric orbits:
$$T{gw}={3\over 85}{c^{5}\over GM}q^{-1}a^{4}(1-e^{2})^{7/2}, \eqno 20$$
where $a={GM\over 2E}$ is the major semi-axis and $e$ is the eccentricity.
Using the standard expressions of celestial mechanics we can express
$T_{gw}$ in terms of variables $x$ and $\tilde E$:
$$T{gw}={3\cdot 4^{7}\over 85\sqrt 2}q^{-1}({GM\over c^{2}r_{a}})
(x^{7}{\tilde E}^{-1/2})\sqrt {{r_{a}^{3}\over GM}}. \eqno 21$$  
As one can see, $T_{gw}$ sharply decreases $\sim x^{7}$
with decrease of $x$.

A star has a considerable probability to be captured on a highly bound
orbit when its angular momentum $x$ is smaller than the size of ``gravitational
wave loss cone'' (see also SR) defined as
$$x_{gw}=\sqrt{x^{2}_{2_wd}{T_{gw}(x_{gw})\over P(E)}}. \eqno 22$$
Using equations (17,19), we obtain
$$x_{gw}={1\over 4}{({85\pi \over 3C_{1}})}^{1/5}{\tilde E}^{1\over 5\beta}
\approx {0.6 \over C_{1}^{1/5}}{\tilde E}^{1\over 5\beta}. \eqno 23$$

Action of two body gravitational encounters and emission of gravitational waves
changes the distribution function $f_{wd}(x, \tilde E)$. To find this change,
one can use the standard Fokker-Plank approach to the problems of stellar dynamics
(e.g. Spitzer 1987 and references therein). However,
the Fokker-Plank equation must be properly modified to take into account emission
of gravitational waves. It is easy to see that the stationary variant of the
modified equation should have the form:
$${\partial \over \partial E}({\dot E_{gw}} {\cal N}_{wd})
+{\partial \over \partial J}({\dot J_{gw}} {\cal N}_{wd})=\hat F_{F-P}({\cal N}), 
\eqno 24$$
Here  
$${\cal N}_{wd}=8\pi^{2}f_{wd}P(E)J \eqno 25$$ 
is the distribution function of white dwarfs over energy and angular momentum.
It is normalized by the condition: $\int dEdJ{\cal N}_{wd}=N_{wd}$.
$\dot E_{gw}(E,J)$ and $\dot J_{gw}(E,J)$ are, respectively, the rate of change of
binding energy and angular momentum of a particular star 
due to emission of gravitational waves. Explicitly, $\dot E_{gw}(E,J)$ and 
$\dot J_{gw}(E,J)$  were calculated by Peters 1964.  $\hat F_{F-P}$ is the standard
Fokker-Plank operator. We need below only the expression for $\dot E_{gw}(E,J)$:
$$\dot E_{gw}={85\over 3\cdot 4^{7}x^{7}}qc^{2}\sqrt {{GM \over r_{a}^{3}}}{\tilde E}^{3/2}.
\eqno 26$$

In general, equation (24) is too complicated to be solved analytically. 
However, we can make
several simplifying assumptions. Namely, we use the standard 
approximations of the loss-cone theory (e.g. LS) 
neglecting differentiation over energy in the Fokker-Plank operator and considering
the diffusion coefficients in the low angular momentum limit. In this limit, 
the right hand side of (24) takes the form:
$$4\pi^2j^{2}_{2_wd}{\partial \over \partial J}J {\partial \over \partial J}f_{wd}.
\eqno 27$$      
Next, we can neglect the term ${\partial \over \partial J}({\dot J_{gw}} {\cal N}_{wd})$
on the left hand side of (24). Indeed, this term is
$\sim E/c^2$ times smaller than the leading term 
${\partial \over \partial E}({\dot E_{gw}} {\cal N}_{wd})$.

After these assumptions being accepted, equation (24) is reduced to a parabolic
type equation
$${\partial \over \partial E}({\dot E_{gw}}P(E){f}_{wd})=
{j^{2}_{2_wd}\over 2J}{\partial \over \partial J}J 
{\partial \over \partial J}f_{wd},
\eqno 28$$
Substituting (19,26) in (28), we obtain
$${85\sqrt 2 \pi \over 3\cdot 4^{5}C_{1}}{\tilde E}^{5/2-p}{\partial \over
\partial \tilde E}f_{wd}=x^{6}{\partial \over \partial x}x {\partial \over \partial x}f_{wd}.
\eqno 29$$      
This equation should be solved with the outer boundary condition
$$f_{wd}(E, J_{max})=\delta CE^{p}, \eqno 30$$
where $J_{max}={GM\over \sqrt {2E}}$ is the angular momentum of a circular
orbit. Instead of $J_{max}$ we use below the corresponding dimensionless angular
momentum:
 $$x_{max}={J_{max}\over J_{crit}}={1\over 4}{({c^{2}r_{a}\over 2GM})}^{1/2}
{\tilde E}^{-1/2}. \eqno 31$$ 
The condition of absorbing wall
$$f_{wd}(x=1)=0 \eqno 32$$ 
can be imposed as an approximate inner boundary condition. 

Equation (29) should be solved separately for the degenerate case $p=0$ and
for the general case $p > 0$. At first let us consider the case $p=0$.
For this case the ``stationary'' solution 
(${\partial \over \partial \tilde E}f_{wd}=0$) is compatible with the boundary conditions
(30) and (32). This solution has the standard form (e.g LS, Cohn $\&$ Kulsrud 1978):
$$f_{wd}=Aln(x), \eqno 33$$
where the normalization constant $A$ can be determined from the outer boundary condition.
The value of the distribution function at $x=x_{max}$:  
$$f_{wd}(x=x_{max})=Aln({J_{max}\over J_{crit}})=
Aln({1\over 4}{({c^{2}r_{a}\over 2GM})}^{1/2}{\tilde E}^{-1/2}) $$
depends logarithmically on energy, and therefore
it cannot be directly equated to (30). Therefore, we equate the last expression and 
(30) at some
fixed energy $E_{*}$ which is specified below (see equation (40)). Similar to the
usual loss cone theory (LS), this leads to slow logarithmic dependence of the 
effective power $p_{eff}\equiv {dln(f_{wd})\over dln(x)}$  of the distribution 
function on energy. By this way we obtain the normalization
constant in the form:
$$A={\delta C\over ln \Lambda_{2}},  
\eqno 34$$
where
$$ln \Lambda_{2}=ln({1\over 4}{({c^{2}r_{a}\over 2GM})}^{1/2}
 E_{*}^{-1/2}). \eqno 35$$
The constant $C$ in (34) is calculated with help of (3),(5):
$$C={2^{-13/2}9\pi^{-2}q^{-1}\over {(GMr_{a})}^{3/2}}. \eqno 36$$
For $p=0$ the flow of white dwarfs toward larger energies
$$\dot N^{0}_{wd}\equiv \int^{J_{max}}_{J_{crit}}dJN_{wd}\dot E_{gw}$$
does not depend on energy. It is calculated with help of (3), (5,6),
(25,26),(33,34):
$$\dot N^{0}_{wd}={51\pi\over 5\cdot 4^{7}}{\delta \over ln \Lambda_{2}}
\sqrt{{GM\over r_{a}^{3}}}. \eqno 37$$
For small energies only a small fraction of white dwarfs can be carried toward
sufficiently large energies by $\dot N_{wd}$, and the main fraction is absorbed
by the absorbing wall at $x=1$. Alternatively, at large energies,
almost all white dwarfs diffused to small angular momenta of order of
$x_{gw}$ form the flow
$\dot N_{wd}$. It is reasonable to identify the boundary between the small and
large energies as the energy $E_{*}$ where the logarithm (35) is evaluated.
To find $E_{*}$ we calculate the diffusion flow per unit of energy:
$${d \dot N_{diff}\over dE}= 4\pi^{2}j^{2}_{2_wd}J{\partial f_{wd}\over \partial J}=
{2^{-9/2}C_{1}\delta \over {(GMr_{a})}^{1/2}}{\tilde E}^{-5/2}. \eqno 38$$
The condition 
$$\int^{\infty}_{E_{*}}d\dot N_{diff}=\dot N^{0}_{wd} \eqno 39$$
gives an implicit equation for $E_{*}$.
This equation has an approximate solution
$$E_{*}\approx 72{({ln \Lambda_{2}C_{1}\over 4.5})}^{2/3}, \eqno 40$$
and 
$$ln \Lambda_{2}\approx 4.5+0.5ln({70\over E_{*}}{r_{1}\over M_{6}}), \eqno 41$$
where $r_{1}=r_{a}/1pc$ and $M_{6}=M/M_{\odot}$.  
It is important to note that the energy $E_{*}$ is of order of the ``collision''
cutoff energy $E_{coll}$ for the normal stars (see eq. (8)), but much smaller than
the ``collision''
cutoff energy $E^{wd}_{coll}$ for the white dwarfs  (eq. (9)). Thus, our assumption of
absence of the normal stars at energies of interest is fulfilled. 
 
Now let us consider the general case $p > 0$.

Before solving equation (29) for $p > 0$, let us point out that       
this equation can be brought into a standard form by the following change of
variables:
$$y=x^{-5/2}, \quad \tau=C_{2}{\tilde E}^{-1/\beta}, \eqno 41$$
where
$$C_{2}\approx {15\cdot 4^{4}C_{1}\beta \over 17\sqrt 2 \pi}\approx 159,7C_{1}\beta. \eqno 42$$ 
We have
$${\partial \over \partial \tau}f_{wd}
+ {1\over y}{\partial \over \partial y}y {\partial \over \partial y}f_{wd}=0. \eqno 43$$
Obviously, equation (43) is the standard diffusion equation written in cylindrical coordinates,
but with negative and constant coefficient of diffusion. 
A formal general solution to (43) can be written as:
$$f_{wd}=A\int^{\infty}_{0}d\lambda D(\lambda)e^{-\lambda \tau}\Psi(y), \eqno 44$$ 
where the function $\Psi (y)$ satisfies the Bessel equation for the zero index Bessel functions
of imaginary argument:
$${1\over y}{d \over dy}y {d \over dy}\Psi -\lambda \Psi=0. \eqno 45$$ 

Now we would like to construct the solution to (43) satisfying the boundary conditions
(30) and (32).
At first, formally assuming that the size of ``gravitational wave loss
cone'' $x_{gw}$  is much larger than unity 
\footnote{
In other words, considering the limit $\tau \rightarrow 0$.}
we look for a self-similar solution to (43).
For this solution the inner boundary condition is reduced to the condition of that 
$f_{gw}$ decreases with decrease of $x$.
The McDonald function $K_{0}(z)$ decreases with increase of the argument, and hence 
the self-similar solution to (43) must be proportional to  $K_{0}(\lambda^{1/2}y)$.
We choose the function $D(\lambda )$
to be
$$D(\lambda )=\lambda ^{\beta p -1}, \eqno 46$$
and obtain:
$$f^{ss}_{wd}=A\int^{\infty}_{0}d\lambda \lambda ^{\beta p -1}
e^{-\lambda \tau}K_{0}(\lambda^{1/2}y).
\eqno 47$$ 
It is easy to see that the solution (47) has a self-similar form indeed. For that let us
change the integration variable in (47): $\tilde \lambda =\lambda \tau $, and introduce
the self-similar coordinate:
$$\xi={y\over \tau^{1/2}}. \eqno 48$$
We have
$$f_{wd}=\tilde A {\tilde E}^{p}\int^{\infty}_{0}d\tilde \lambda
{\tilde \lambda }^{\beta p -1}
e^{-\tilde \lambda}K_{0}({\tilde \lambda}^{1/2}\xi),
\eqno 49$$
where $\tilde A={A\over C_{2}^{\beta p}}$.  
The integral in (49) is proportional to the Whittaker function
$W_{\mu,0}(z)$:
$$\int^{\infty}_{0}d\tilde \lambda
{\tilde \lambda }^{\beta p -1}
e^{-\tilde \lambda}K_{0}({\tilde \lambda}^{1/2}\xi)=\Gamma^{2}(\beta p){e^{\xi^{2}/8}
\over \xi}W_{{1\over 2}-\beta p}({\xi^{2}\over 4}). \eqno 50$$
It is important to note that the self-similar coordinate $\xi$ defines a scale
in the angular momentum space of order of $x_{gw}$:
$$\xi={({x\over x_{*}})}^{-5/2}, $$
where
$$x_{*}={\tau}^{-1/5}={1\over 4}{({68\sqrt 2\pi\over 15C_{1}\beta})}^{1/5}
{\tilde E}^{1\over 5\beta}\approx {0.46\over {(C_{1}\beta)}^{1/5}}
{\tilde E}^{1\over 5\beta}. \eqno 51$$

Let us discuss the asymptotic behavior of (49). For large $\xi$ (small $x$) we use
the approximate asymptotic expression for the Whittaker function
$$W_{\mu, 0}(z)\approx z^{\mu}e^{-z/2}, \eqno 52$$  
to obtain
$$f^{ss}_{wd}\approx {1\over 2}{(4C_{2})}^{\gamma}\tilde A \Gamma^{2} (\gamma)x^{5\gamma},
\eqno 53$$
where $\gamma =\beta p$.
As it is seen from (53), the function $f^{ss}_{wd}$ has a constant nonzero value
at $x=1$. This contradicts to the inner boundary condition (32). However, we can
subtract this value from $f^{ss}_{wd}$ and satisfy the boundary condition.
We have:
$$f_{wd}=f_{wd}^{ss}-f^{ss}_{wd}(x=1)=\tilde A\Gamma^{2}(\gamma )\times $$
$$\times \lbrace {\tilde E}^{p}{e^{\xi^{2}/8}\over \xi}W_{1/2-\gamma,0}({\xi^{2}\over 4})-
{{(4C_{2})}^{\gamma}\over 2}\rbrace. \eqno 54$$  
In the limit of small $\xi$ the function (54) has the form (see Appendix):
$$f_{wd}\approx \tilde A\Gamma (\gamma ){\tilde E}^{p}\times $$
$$\times \lbrace
ln(2\xi^{-1})-(C_{e}+\psi(\gamma )/2) - \Gamma (\gamma )
{{(4C_{2})}^{\gamma }\over 2}{\tilde E}^{-p} \rbrace, \eqno 55$$
where $C_{e}\approx 0.577$ is the Euler constant, and 
$\psi (z)={d\over dz}ln\Gamma(z)$. The expression (55) is used 
for normalization of the distribution function (54).
Similar to the case $p=0$, $f_{wd}(x=x_{max})$ is not an exact power law
function of energy, and we should find the normalization constant
$\tilde A$ at some fixed energy $E_{*}$. It is reasonable to identify 
this energy as the energy corresponding to the ``time'' $\tau =1$
(see equation (41)):
$$E_{*}=C_{2}^{\beta}. \eqno 56$$
Taking into account (41), (48) and (55), we obtain   
$$f_{wd}(x=x_{max})(\tilde E =E_{*})  =\tilde A \Gamma (\gamma ){\tilde E}^{p}\times $$
$$\times \lbrace ln(2 x_{max}^{5/2})-(C_{e}+\psi (\gamma )/2 +4^{\gamma }\Gamma (\gamma )/2)\rbrace.
\eqno 57 $$
The last term in the brackets  
could be important if only $\gamma \rightarrow 0$, where we approximate
$$\psi (\gamma )/2 +4^{\gamma }\Gamma (\gamma )/2 \approx ln(4). \eqno 57$$
Using equations (30,31) and (56,57), we obtain the normalization constant
in the form
$$\tilde A ={\delta C \over \Gamma (\gamma ) ln \Lambda_{3}}{({GM \over r_{a}})}^{p}, \eqno 58$$ 
where
$$ln \Lambda_{3}=ln ({e^{-C_{e}}\over 32}{({c^{2}r_{a}\over 2GM})}^{5/4}{\tilde E_{*}}^{-5/4})
\approx $$
$$11 +ln \lbrace {({r_{1}\over M_{6}})}^{5/4}C_{2}^{\beta (1/4 -p)} \rbrace . \eqno 59$$ 
Note, that equations (55) and (58) tell that the ``effective power'' of the distribution
$f_{wd}$ differs from the parameter $p$ on value of order of 0.1.   
It is also important to point out that the solution (54) satisfies the boundary conditions
in the limit $\tau \rightarrow 0$
(e.g $\tilde E \gg E_{*}$). We assume below that the solution (54) also 
gives a reasonable approximation
for the moderate values of $\tilde E$: $\tilde E \ge E_{*}$. 
For $\tau > 1$ ($\tilde E < E_{*}$) the effects of gravitational
radiations are negligible and we can use the standard solution (33). 

Now let us calculate the capture rate of white dwarfs 
for the case $p > 0$. At first it is convenient to calculate  
the capture rate of white dwarfs due to emission of gravitational waves
per unit of energy:
$${d\dot N_{wd}\over dE} \equiv\int^{J_{max}}_{J_{crit}}dJ {\partial \over \partial E}(N_{wd}\dot E_{gw})
\eqno 60$$
This quantity can be expressed as the difference between the diffusion flow of the stars
at large values of $x$ and the diffusion flow at $x=1$:
$${d\dot N_{wd}\over dE}={d\dot N_{diff}\over dE}|_{x\rightarrow x_{max}}
-{d\dot N_{diff}\over dE}|_{x=1}, \eqno 61 $$
where  
$${d\dot N_{diff}\over dE}=4\pi^{2}j^{2}_{2_wd}J{\partial \over \partial J} f_{wd}. \eqno 62$$
We calculate the diffusion flows with help of (13), (53,55):
$${d\dot N_{diff}\over dE}|_{x\rightarrow x_{max}}=
10\pi^{2}j^{2}_{2_wd}{\delta C {\tilde E}^{p}\over ln \Lambda_{3}}{({GM \over r_{a}})}^{p} , \eqno 63$$
and
$${d\dot N_{diff}\over dE}|_{x=1}=10\pi^{2}j^{2}_{2_wd}\gamma 4^{\gamma}\Gamma (\gamma )
{\delta C \over ln \Lambda_{3}}{({GM \over r_{a}})}^{p}. \eqno 64$$
Note that the diffusion flow at large $x$ (equation 63) is close to the similar expression obtained 
by LS in frameworks of ``one dimensional'' loss cone theory.
Substituting (63), (64) in (61), and using (3-5), we obtain:
$$ {d\dot N_{wd}\over dE}=2^{-7/2}{5\delta C_{1}\over \beta B(3/2;p+1)ln \Lambda_{3}}\times $$
$$ \times \lbrace {\tilde E}^{2p-5/2}
-\gamma 4^{\gamma}\Gamma (\gamma )E^{p}_{*}{\tilde E}^{p-5/2}\rbrace {1\over {(GMr_{a})}^{1/2}}   
\eqno 65 $$  
The total capture rate of white dwarfs can be expressed as
$$\dot N_{wd} \approx \int^{E^{wd}_{coll}}_{E_{1}}d\dot N_{wd} +\dot N^{0}_{wd}, \eqno 66$$ 
where we use equation (37) for $\dot N^{0}_{wd}$
and the energy scale
$$E_{1}={(\gamma 4^{\gamma}\Gamma (\gamma ))}^{1/p}E_{*} \eqno 67$$
is defined by the condition ${d\dot N_{wd}\over dE}(E_{1})=0$.
Explicitly integrating (66) and using (42), (56), (67) we have: 
$$\dot N_{wd}\approx {17\pi \over 6\cdot 4^{6}}({p-(3/2-p)\epsilon^{3/2-2p}+
(3/2-2p)\epsilon^{3/2-p}\over 3/2-2p})\times $$
$$\times {{(\gamma \Gamma (\gamma ))}^{-1/\gamma} 
\delta \over \beta B(3/2;p+1) ln \Lambda_{3}}E_{1}^{p} 
\sqrt {{GM\over r_{a}^{3}}} +\dot N_{wd}^{0}, \eqno 68$$
where $\epsilon={E_{1}\over \tilde E^{wd}_{coll}} \le 1$.
This expression can be rewritten as
$$\dot N_{wd}\approx K(p)\dot N_{wd}^{0}, \eqno 69$$
where
$$K(p)=1+{20\over 18}({p-(3/2-p)\epsilon^{3/2-2p}+
(3/2-2p)\epsilon^{3/2-p}\over 3/2-2p})\times $$
$$\times {{(\gamma \Gamma (\gamma ))}^{-1/\gamma} 
ln \Lambda_{2} \over \beta B(3/2;p+1) ln \Lambda_{3}}E_{1}^{p}. \eqno 70$$ 
The function $K(p)$ is plotted in Fig. (1) for $ln \Lambda =ln(0.5\cdot 10^{5})\approx 11$,
$ln \Lambda_{2}=4.5$, $ln \Lambda_{3}=11$, 
and $\delta=0.1$.
\begin{figure}
\vspace{8cm}\includegraphics{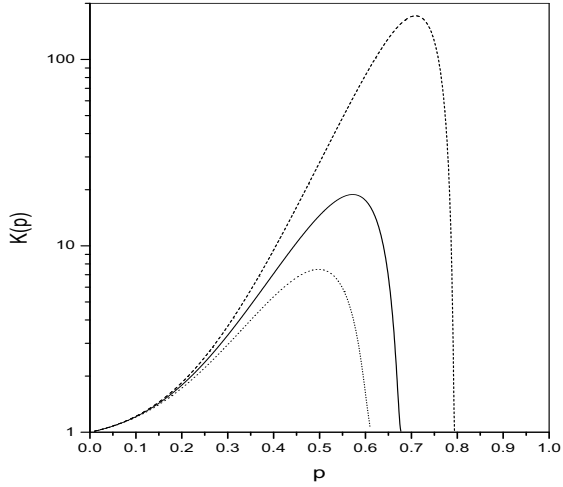}
\caption{ 
The ratio $K(p)={\dot N_{wd}\over \dot N^{0}_{wd}}$ is shown as a function of $p$.
Solid curve corresponds to $q_{6}r_{1}=1$ and $E_{coll}^{wd}=4300$. Dashed and
dotted curves correspond to $q_{6}r_{1}=5$ ( $E_{coll}^{wd}=21500$) and 
$q_{6}r_{1}=0.5$ ($E_{coll}^{wd}=2150$), respectively.
The solid curve  has a maximum at $p\approx 0.57$.}
\end{figure}

One can see from this Fig. that the capture rate depends non-monotonically
on $p$, it has a  maximum at $p\approx 0.57$ ($K(p=0.57)\sim 19$), for larger values of  
$p$ our assumption of absence of the white dwarfs at energies larger than $E_{coll}^{wd}$
leads to suppression of the capture rate (see also Fig. 2).
The capture rate is mainly determined by stars with ``initial'' energy
$$E_{in}={({3/2-p\over 3/2 -2p})}^{1/p}E_{1}, \eqno 71$$
defined by the condition ${d\over dE}({d\dot N_{wd}\over d ln E}(E_{in}))=0$
and these stars have their ``initial'' angular momentum
of order of
$$x_{in}=x_{*}(E_{in}). \eqno 72$$
We show the dependence of $E_{in}$ and $x_{in}$ on $p$ in Fig. 2 and Fig 3, 
respectively.
\begin{figure}
\vspace{8cm}\includegraphics{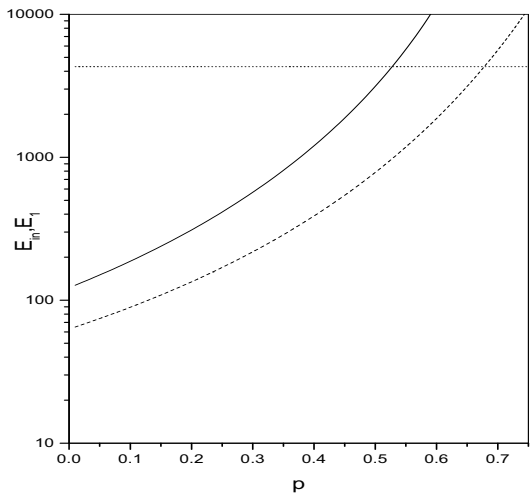}
\caption{
The characteristic energies $E_{in}$ (solid curve) and $E_{1}$ (dashed curve)
are shown as functions of $p$. They
monotonically increase with
$p$. For $p > 0.53$ $E_{in}$ is larger than $E^{wd}_{coll}=4300$ (dotted line),
and for $p > 0.675$ $E_{1}$ is larger than  $E^{wd}_{coll}$.}
\end{figure}
\begin{figure}
\vspace{8cm}\includegraphics{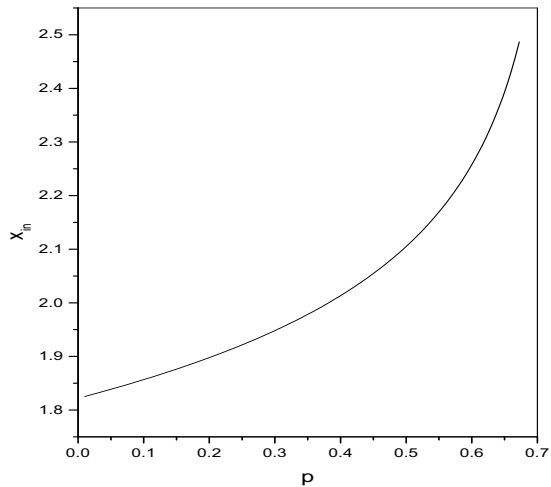}
\caption{ 
The characteristic angular momentum $x_{in}$ is shown as a function of $p$. It has
the value $\approx 1.8$ for $p=0$ and monotonically increases with $p$. For
specially important $p=1/4$ $x_{in}\approx 2$. 
}
\end{figure}

It is instructive to compare the capture rate obtained in this paper with
that was obtained by SR. The 'diffusion' capture rate of SR can be written as
\footnote{To obtain this equation we use eq. (15) of SR with $\beta=5$ and
assume that the inner part of the cusp consists purely of white dwarfs.}
$$\dot N^{SR}(r)\approx 5{({N_{wd}(r)m_{wd}\over M})}^{8/5}({r_{s}\over r})P^{-1},$$
where $r_{s}={2GM\over c^{2}}$, and $N_{wd}$ is the number of white dwarfs within
the sphere of radius $r$. In order to describe the spatial scale SR use the
radial coordinate $r$ while we work with the energy coordinate $E$. Since the
stars spend most of their time near their apocenters $r_{ap}=GM/E$, it is reasonable
to use the following rule of change of variables: $r=r_{ap}$.
Using equations (4,19) and definition of $\delta$, we have
$$\dot N^{SR}(r)\approx {5\sqrt 2\over \pi}\delta^{8/5}({r_{s}\over r_{a}})
{({r\over r_{a}})}^{-(1/10+8p/5)}\sqrt{{GM\over r_{a}^{3}} }=
{5\sqrt 2\over \pi}\delta^{8/5}({r_{s}\over r_{a}})
{\tilde E}^{(1/10+8p/5)}\sqrt{{GM\over r_{a}^{3}}}.$$  
One can see from this equation that $\dot N^{SR}$ grows with energy,
and therefore we use the 'cutoff' energy $E^{wd}_{coll}$ (equation 9)
to obtain the total rate 
$$\dot N^{SR}\approx 6\cdot 10^{-9}{M_{6}\over r_{1}}\delta_{1}^{8/5}
{(\tilde E^{wd}_{coll})}^{8p/5+1/10}\sqrt{GM\over r_{a}^{3}}, \eqno 73$$
where $\delta_{1}=10\delta $.
We show the ratio $R=\dot N^{SR}/\dot N_{wd}$ in Fig 4.
\begin{figure}
\vspace{8cm}\includegraphics{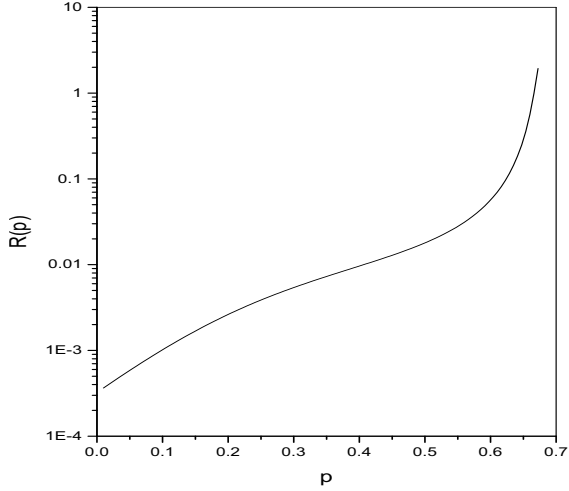}
\caption{ 
The  ratio $R=\dot N^{SR}/\dot N_{wd}$ is shown as a function of $p$.}
\end{figure}
One can see that the capture rate of SR is much smaller than
our capture rate for small values of $p$ 
($\dot N^{SR}(p=0)/\dot N_{wd}(p=0)\approx 3\cdot 10^{-4}$ and  
$\dot N^{SR}(p=1/4)/\dot N_{wd}(p=1/4) \approx 10^{-2}$)
\footnote{
Technically, SR made a mistake in derivation of their capture rate.
In their equation (14) they used the so-called ``relaxation time-scale''
in the denominator instead of correct time-scale of orbit decay
due to gravitational radiation, which is given by our equation (21).}.      
Here we would like to note that SR also obtain the capture rate due to
large angle deflections of stars which is several times larger (for the case 
$p=0$) than our capture rate. The large angle deflections of the stars
cannot be treated in framework of the Fokker-Plank approach and are
beyond of the scope of this paper.

Finally, let us represent the capture rate in astrophysical units:
$$\dot N_{wd}\approx 3\cdot 10^{-9}K(p)\delta_{1}
({4.5\over ln \Lambda_{2}}){M_{6}^{1/2}\over r_{1}^{3/2}}yr^{-1},
\eqno 74$$
This expression can be directly compared with Monte Carlo simulations
made by Hils and Bender 1995 for $p=0$. Hils and Bender obtained the
capture rate $\dot N^{HB}_{wd}\approx 1.9\cdot 10^{-8}yr^{-1}$
which is $6.3$ times larger than our value. Unfortunately,
Hils and Bender did not publish details of their method,
and analysis of this disagreement is not possible.  

\section{Probability of finding of the source of gravitational waves}

The probability ${\it Pr}$ of finding of a galaxy with a white dwarf orbiting around central black hole
and emitting gravitational radiation in a detectable amount can be defined as 
$${\it Pr}=T^{f}_{gw}\dot N^{wd}, \eqno  75 $$ 
where $T^{f}_{gw}$ is the decay time of an orbit of a white dwarf which is sufficiently close
to the black hole to yield a significant amplitude of gravitational waves. Clearly,
this time depends not only on physical conditions inside the galactic cusps, but also
on properties of gravitational wave antenna receiving the signal, in particular, on
frequency band available for such antenna. Therefore, we assign the superscript $f$ to
$T^{f}_{gw}$. In particular, the maximal sensitivity of LISA gravitational wave antenna
will be in the frequency range $10^{-3}Hz \le f \le 10^{-1}Hz$, and sharply decreases with
decreasing frequency
\footnote{
For an overview of LISA mission see
http://lisa.jpl.nasa.gov.}. LISA would be able to detect a gravitational wave with amplitude $h$ of
the order $\sim 10^{-23}$ at $f\sim 10^{-3}Hz$ with      
a signal-to-noise ratio $\sim 5$. For moderately eccentric orbits the frequency
of gravitational waves is inversely proportional to the period $f \sim 1/P(E)$, and therefore
we are interested in periods $P(E)$ of order of $10^{3}s$. For such periods and black hole masses,
$\sim 10^{6}M_{\odot}$ the semi-major axis $a$ of the orbit is close to five gravitational radii:
$$a=10{({P_{3}\over M_{6}})}^{2/3}{GM\over c^{2}}, \eqno 76$$ 
where $P_{3}={P(E)\over 10^{3}s}$.
In such situation, strictly speaking, one 
should use relativistic expressions for $T^{f}_{gw}$ and other quantities of interest.
However, the relativistic expressions are rather complicated, and we are going to use 
expressions calculated in the Newtonian approximation assuming that this
would not significantly alter our order-of-magnitude estimates.

Now we would like to discuss the following problem. Suppose that white dwarfs are supplied
on highly eccentric orbits with the rate given by equation (69) and 
with energy and angular momentum given by equations (71,72). After that 
the orbital parameters are 
changed only due to emission of gravitational radiation. What is the eccentricity of 
the orbit when its period is of order of $10^{3}s$? This problem can be solved with help
of relation between parameters of the orbit evolving due to emission of
gravitational radiation (Peters 1964):
$$x(e)\approx 0.94x_{in}e^{6/19}H(e), \eqno 77$$   
where
$$H(e)={(1+{121\over 304}e^{2})}^{435/2299}$$
slowly changes with $e$.
We can rewrite equation (77) as
$$a\approx 14.8 {GM\over c^{2}}x_{in}^{2}e^{12/19}H^{2}(e), \eqno 78$$
substitute (76) into (78), and obtain an implicit equation for the eccentricity $e$. Solution
of this equation depends parametrically on $p$ (through the dependence 
of $x_{in}$ on $p$, equation (72)), and also on the ratio $({P_{3}\over M_{6}})$. This 
solution is shown in Fig. 5. 
\begin{figure}
\vspace{8cm}\includegraphics{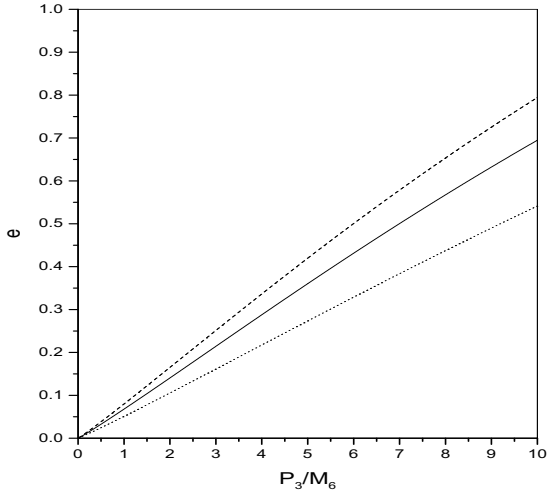}
\caption{
Dependence of the eccentricity $e$ on the ratio $P_{3}/M_{6}$. Solid curve corresponds to $p=1/4$,
dashed and dotted curves correspond to $p=0$, $p=0.5$, respectively. Note that these curves are
close to straight lines.}
\end{figure}
As it is seen from this Fig., for the most
interesting case $({P_{3}\over M_{6}})\sim 1$, the eccentricity is small ($e\le 0.1$).
\footnote   
{Note, however that for $({P_{3}\over M_{6}})=10$ the eccentricity is substantial $e\le 0.8$.}
Therefore, we neglect the dependence of the decay time $T^{f}_{gw}$ on the eccentricity
and use $T^{f}_{gw}$ calculated by Peters 1964 for circular orbits:
$$T^{f}_{gw}={5\over 256}q^{-1}{({ac^{2}\over GM})}^{4}{GM\over c^{3}}\approx
31M_{6}{({P_{3}\over M_{6}})}^{8/3}yr. \eqno 79$$
Substituting (74) and (79) into equation (75) we finally have:
$${\it Pr}\approx 10^{-7}K(p)\delta_{1}
({4.5\over ln \Lambda_{2}}){({M_{6}\over r_{1}})}^{3/2}{({P_{3}\over M_{6}})}^{8/3}. \eqno 80$$
Note that ${\it Pr}$ strongly depends on the ratio $ P_{3}/M_{6}$.

\section{Conclusions}

We calculate the formation rate of a close binary in galactic cusps 
(the capture rate) consisting
of a super-massive black hole and a compact object (presumably a white dwarf).
We also calculate the probability ${\it Pr}$ to find a galaxy with such a binary
emitting gravitational radiation at frequency $f\sim 10^{-3}Hz$.
We take into account two main process determining the capture rate: 1) two-body 
gravitational encounters of the compact object, 2) energy loss due to
emission of gravitational radiation, and calculate the capture rate analytically,
in frameworks of the so-called loss cone approximation. Our main results are given
by equations (37),(69) and (80). These equations explicitly show the dependence of
the capture rate and probability ${\it Pr}$ on main parameters of galactic cusps:
their radii, the relative number of compact objects in them, the ``sharpness'' $p$ of
increase of the number density of compact objects toward the center of a galaxy and
the frequency band of gravitational wave antenna receiving the signal.
They could be easily used for estimates of the capture rate and probability ${\it Pr}$
for a broad range of the main parameters. We correct a mistake made by previous 
researchers in estimate of the capture rate.
Our results are obtained in assumption of power law cusps with density 
$n \sim r^{-(3/2+p)}$. However, generalization of our results on more realistic
cusp profiles (say, with different powers $p$ in the outer and inner regions
of the cusp, e.g Alexander 1999) is straightforward.

Unfortunately, not so much is known about galactic cusps from observations.
Therefore, our results should be considered as qualitative only. Also, much more
robust estimates could be obtained using modern computational methods. However,
our simple analytical approach clearly shows the dependence on parameters, and
therefore, could be used as a guidance for choice of parameters for numerical
studies.   

In their recent review of black holes found in the galactic centers, Kormendy
and Gebhardt 2001 identify three black hole with masses of order of 
$\sim 10^{6} M_{\odot}$ (one of them is in our own Galaxy) with distance
smaller than $13,2Mpc$. Therefore, it seems reasonable to estimate
the density of potential sources of gravitational radiation as 
$\ge 10^{-3}Mpc^{-3}$
\footnote{
This density could be larger due to selection effects. On the other hand,
the probability ${\it Pr}$ is sensitive to the size $r_{a}$ of the cusp (eq. (80)).
If all potential sources have the cusps with sizes $\gg 1pc$, the probability
would be suppressed.}. 
Assuming that we need $\sim 10^{7}$ potential
sources to observe at least one event (equation (80)), we need
the sensitivity of gravitational wave antenna to be sufficient to detect a source at
the distance $D \ge 10^{3}Mpc$. The dimensionless amplitude of a gravitational
wave coming from a white dwarf orbiting around a black hole
with period $\sim 10^{3}s$ at the distance $\sim 10^{3}Mpc$ from an observer
is of order of $10^{-23}$. This amplitude could be easily detected by
a gravitational wave antenna of LISA type. Therefore, we could optimistically
expect several events during the life-time of LISA  mission $\sim 3 yr$. 
Relativistic effects (e.g. influence of Einstein precession and 
Lense-Thirring precession 
for a rotating black hole on the shape of the signal) may 
help to distinguish these events from other
sources of gravitational radiation.

\section*{Acknowledgments}
I am indebted to N. Kardashev who attracted my attention to this problem. 
I also thank M. Prokhorov and the referee
for useful remarks, A. Illarionov,
D. Kompaneets, A. Rodin and A. Polnarev for discussions. 
This work was supported in part by RFBR grant 00-02-16135.

\appendix
\section[]{}

It is rather difficult to find decomposition of the Whittaker function $W_{\mu,0}(z)$
for small values of $z$ in the standard reference books on special functions. 
Therefore, we write below the formulae which are used to obtain (55) from (54).

We use decomposition of the McDonald function for small $z$:
$$K_{0}(z)\approx ln({2\over z})-C_{e}, \eqno A1$$
the representation of the Gamma function:
$$\Gamma (z)=\int^{\infty}_{0}d\lambda \lambda^{z-1}e^{-\lambda}, \eqno A3$$
and the consequence of (A3):
$$\psi(z)\equiv {d\over dz}ln \Gamma(z)={1\over \Gamma (z)}
\int^{\infty}_{0}d\lambda \lambda^{z-1}e^{-\lambda}ln \lambda. \eqno A3$$

\bsp

\label{lastpage}


\begin{thebibliography}{99}
\bibitem{b1} Alexander T., 1999, ApJ, 527, 835 
\bibitem{b2} Bahcall J. N., Wolf R. A., 1976, ApJ, 209, 214
\bibitem{b3} Bisnovatyi-Kogan G. S., Churaev R. S., Kolosov B. I., 1982,
A$\&$A, 113, 179
\bibitem{b4} Chandrasekhar, S., 1942, Principles of Stellar Dynamics, Chicago,
University of Chicago Press
\bibitem{b5} Cohn H., Kulsrud R. M., 1978, ApJ, 226, 1087
\bibitem{b6} Frolov V. P., Novikov I. D., 1998, Black hole physics: basic concepts
and new developments, Dordrecht: Kluwer Academic
\bibitem{b7}  Gradshtein I. S., Ryzhik I. M., 1994, Table of Integrals, Series and Products,
Academic Press
\bibitem{b8} Gurevich A. V., Geomagnetism and Aeronomy, 1964, 4, 192
\bibitem{b9} Hils D., Bender P. L., 1995, ApJ, 445, L7
\bibitem{b10} Kormendy J., Gebhardt K., 2001, in: 
The 20th Texas Symposium on Relativistic Astrophysics, ed. H. Martel and 
J. C. Wheeler, AIP 
\bibitem{b11} Lightman A. P., Shapiro S. L., 1977, ApJ, 211, 244 
\bibitem{b12} Peters P. C., 1964, 136, 1224
\bibitem{b13} Peebles, P. J. E., 1972, ApJ, 178, 371 
\bibitem{b14} Rosenbluth M. N., MacDonald W. M., Judd D. L., 1957,
Rep. Mod. Phys., 107, 1 
\bibitem{b15} Shapiro, S. L., Marchant, A. B., 1978, ApJ, 225, 603
\bibitem{b16} Sigurdsson S., Rees M. J., 1997, MNRAS, 284, 318
\bibitem{b17} Spitzer L., 1987, Dynamical evolution of globular clusters,
Princeton, NJ, Princeton University Press
\bibitem{b18} Young P., 1980, ApJ, 249, 1232


\end{thebibliography}
\end{document}